\documentclass[11pt, a4paper]{scrartcl}
\usepackage[numbers]{natbib}
\usepackage{booktabs}
\usepackage{amsmath, amssymb}
\usepackage[text={160.5mm,226mm},%
papersize={210mm,280mm},%
columnsep=12pt,%
headsep=21pt,%
centering]{geometry}

\usepackage{listings}
\usepackage{comment}
\usepackage{color}  
\definecolor{arsenic}{rgb}{0.23, 0.27, 0.29}
\usepackage{moreverb,url}
\usepackage{Sweave}
\usepackage[colorlinks,bookmarksopen,bookmarksnumbered,citecolor=red,urlcolor=red]{hyperref}

\newcommand\BibTeX{{\rmfamily B\kern-.05em \textsc{i\kern-.025em b}\kern-.08em
T\kern-.1667em\lower.7ex\hbox{E}\kern-.125emX}}

\begin{document}
\begin{center}
{\LARGE{\bf An intuitive Bayesian spatial model for disease mapping that accounts for scaling
}}
\end{center}
\vskip 2mm
\begin{center}
Andrea Riebler \footnote{(\baselineskip=10pt to whom correspondence should be addressed) Department of Mathematical Sciences, Norwegian University of Science and Technology, Norway,
andrea.riebler@math.ntnu.no},
Sigrunn H.~S{\o}rbye \footnote{\baselineskip=10pt Department of Mathematics and Statistics, UiT The Arctic University of Norway, Norway.},
\setcounter{footnote}{2}
Daniel Simpson
\footnote{\baselineskip=10pt Department of Mathematical Sciences,
  University of Bath, United Kingdom.},
H{\aa}vard Rue$^1$,
\end{center}
%
%
%
{\large{\bf Abstract}}\\
{\small
In recent years, disease mapping studies have become a routine
application within geographical epidemiology and are typically
analysed within a Bayesian hierarchical model formulation. A variety
of model formulations for the latent level have been proposed but all come with
inherent issues. In the classical BYM (Besag, York and Molli{\'e})
model, the spatially structured component cannot be seen independently
from the unstructured component.  This makes prior definitions for
the hyperparameters of the two random effects challenging. There are
alternative model formulations that address this confounding, however,
the issue on how to choose interpretable hyperpriors 
is still unsolved. Here, we discuss a recently proposed parameterisation of the BYM model that leads to
improved parameter control as the hyperparameters can be seen
independently from each other. Furthermore, the need for a
scaled spatial component is addressed,  which facilitates assignment of interpretable 
hyperpriors and make these transferable between 
spatial applications with different graph structures. The hyperparameters themselves are used to define flexible extensions of simple base
models. Consequently, penalised complexity (PC) priors for these parameters can be derived based on the
information-theoretic distance from the flexible model to the base model, giving priors with clear interpretation. We provide implementation details
for the new model formulation which preserve sparsity properties, and we
investigate systematically the model performance and compare it to existing 
parameterisations. Through a simulation study, we show
that the new model performs well, both showing good learning abilities and good shrinkage behaviour.  In terms of model choice criteria, the proposed model 
performs at least equally well as existing
parameterisations, but only the 
new formulation offers parameters that are interpretable and
hyperpriors that have a clear meaning.}
\baselineskip=12pt

%
%
%

\baselineskip=12pt
\par\vspace{0.5cm}\noindent

\section{Introduction}

Over the recent years, there has been much interest in spatial modeling and mapping of
disease or mortality rates. Due to inherent sampling
variability it is not recommended to inspect crude rates directly, but
borrow strength over neighbouring regions to get more reliable
region-specific estimates. The state of the art is to
use Bayesian hierarchical models, where the risk surface is modelled using a set of spatial random
effects, in addition to potentially available covariate
information \cite{lawson-book}. The random effects shall
capture unobserved heterogeneity or spatial correlation that cannot be
explained by the available covariates \cite{lee-etal-2011}. However, there
has been much confusion on the design of the spatial random effects. Besag et
al.\cite{besag-etal-1991} proposed an intrinsic autoregressive model, often referred to as  
the CAR prior or Besag model,  where the spatial effect of a particular region depends 
on the effects of all neighbouring regions. They also proposed the commonly
known BYM model, where an additional unstructured spatial random effect is
included to account for independent region-specific noise. 
 Consequently, various model
modifications and alternative approaches have been proposed, see for
example Leroux et al. 
\cite{leroux-etal-2000}, Stern and Cressie \cite{stern-cressie-2000}
and Dean et al. \cite{dean-etal-2001} The appropriateness and behaviour
of different latent spatial models have been compared in a full
Bayesian context \citep{best-etal-2005,wakefield-2007,macnab-2010,lee-etal-2011},
an empirical
Bayes setting \cite{lee-durban} or using penalized quasi likelihood
\cite{leroux-etal-2000}. It is accepted that the Besag model may lead
to misleading results in the case where there is no spatial correlation
in the data \citep{leroux-etal-2000,wakefield-2007}. For this
purpose, most alternative models propose to account for unstructured
variability in space.

By including both structured and unstructured components in the model,
potential confounding must be addressed
\cite{bernardinelli-etal-1995}, as otherwise it is not clear how to
split the variability over the effects. This problem has motivated development of reparameterised models, in which the precision
parameters of the two components are replaced by a common precision parameter and a mixing parameter, which distribute the
variability between the structured and unstructured components \citep{leroux-etal-2000,dean-etal-2001}.

However, existing approaches do have two issues. First, the spatially
structured component is not scaled. This implies that the precision
parameter does not represent the marginal precision but is confounded
with the mixing parameter. Thus, the effect of any prior assigned to the precision
parameter depends on the graph structure of the application. 
This has the additional effect that a given prior is not transferable
between different applications if the underlying graph changes
\cite{sorbye-rue-2014}. Second, the choice of hyperpriors for the
random effects is not straightforward.  Approaches to design epidemiologically sensible hyperpriors include, among others, Bernardinelli et al. \cite{bernardinelli-etal-1995} and Wakefield
\cite{wakefield-2007}. 

Recently, Simpson et al. \citep{simpson-etal-2015} proposed a new  BYM
parameterisation that accounts for scaling and provides an intuitive way to
define priors by taking the model structure into account. This new
model provides a new way to look at the BYM model. The primary goal is not to
optimize model choice criteria, such as DIC values, but to provide
a sensible model formulation where all parameters have a clear meaning.
The model structure is similar to the Dean
model\cite{dean-etal-2001}, with the crucial modification that
the precision parameter is mapped to the marginal standard deviation. This makes
the parameters of the model interpretable and facilitates assignment of
meaningful hyperpriors. The framework of
penalised complexity (PC) priors is applied to formulate prior
distributions for the hyperparameters.  The spatial model is thereby seen as a 
flexible extension of  two simpler base models that it will shrink towards, if not indicated otherwise by the data
\cite{simpson-etal-2015}. The upper level base model assumes a
constant risk surface, while the lower level model assumes a varying
risk surface over space without spatial autocorrelation. In this paper
we investigate systematically the model behaviour under different
simulation scenarios. Furthermore, we compare this new model
formulation to commonly used model formulations in disease mapping. We point out
differences, focusing on parameter interpretability and
prior distribution assignment. For completeness we also compare
commonly used model choice criteria.

We have chosen to implement all models using integrated nested
Laplace approximations (INLA) \cite{rue-etal-2009}, available to the
applied user via the R-package {\tt INLA} (see \url{www.r-inla.org}). INLA
is straightforward to use for full Bayesian inference in
disease mapping and avoids any Markov chain Monte Carlos techniques
\citep{schrodle-held-2011,schrodle-held-2011b}.

The plan for the paper is as follows. Section~\ref{sec:diseaseMapping}
gives an introduction to disease mapping and motivates the use of Bayesian
hierarchical models. Commonly used spatial models for the latent level
are reviewed in Section~\ref{sec:spatialModels}, before the need of scaling
and consequently the new modified spatial model is
presented. Section~\ref{sec:pcprior} uses the PC-prior framework to
design sensible hyperpriors for the precision and mixing parameter. In
Section~\ref{sec:application},  we investigate the properties and
behaviour of the new 
model in a simulation study, including comparisons to commonly
used models. This section also contains an application to insulin-dependent
diabetes mellitus data in Sardinia. Discussion and concluding
remarks  are given in Section~\ref{sec:discussion}.

\section{Disease Mapping}\label{sec:diseaseMapping}

Disease mapping has a long history in epidemiology and one major
goal is to investigate the geographical distribution of disease
burden \cite{wakefield-2007}. In the following, we assume that our region
of interest is divided into $n$ non-overlapping areas. Let $E_{i}$
denote the number of persons at risk in area $i$ ($i=1, \ldots,
n$). These expected numbers are commonly calculated based on the size
and demographic characteristics of the population living in area $i$
\cite{lee-etal-2011}. Further, let
$y_{i}$ denote the number of cases or deaths in region $i$. When the
disease is non-contagious and rare, it is usually reasonable to assume that
\begin{gather*}
   y_i \mid \theta_i \sim \text{Poisson}(E_i \theta_i),
\end{gather*}
where $\theta_i$, $i=1, \ldots, n$, denotes the underlying true area-specific relative risk \cite{bernardinelli-etal-1995}.

The maximum likelihood estimator of $\theta_i$ is given by $\hat{\theta}_i =
y_i/E_i$ and corresponds to the standardised mortality ratio (SMR) if the
counts represent deaths. Mapping the SMRs directly can, however, be
misleading. Areas with extreme SMR values often have small expected
numbers, so that sampling variability is more pronounced. This is
reflected in large values for $\text{Var}(\hat{\theta}_i) =
\theta_i/E_i$, see Bernardinelli et
al. \cite{bernadinelli-etal-1997} or Wakefield \cite{wakefield-2007} for more
details.

Due to these potential instabilities, methods have been proposed to
smooth the raw estimates over space. Bayesian hierarchical models are
commonly used in this context, where the latent model involves random
effects that borrow strength over neighbouring regions and such
achieve local smoothness \cite{bernadinelli-etal-1997}. We refer to
Lawson \cite{lawson-book} and Banerjee et al. \cite{banerjee-book} for a
detailed description of the class of Bayesian hierarchical models.

A general model formulation is to assume that the log risk $\eta_i =
\log(\theta_i)$ has the decomposition:
\begin{gather*}
  \eta_i = \mu + \mathbf{z}_i^\top \mathbf{\beta} + b_i.
\end{gather*}
Here, $\mu$ denotes the overall risk level, $\mathbf{z}^\top_i =
(z_{i1}, \ldots, z_{ip})^\top$ a set of $p$ covariates with
corresponding regression parameters $\beta=(\beta_1, \ldots,
\beta_p)^\top$, and $b_i$ a random effect. For $\mu$ and
$\mathbf{\beta}$ we assign weekly informative Gaussian distributions
with mean zero and variance 100. The random effects $\mathbf{b} =
  (b_1, \ldots,b_n)^\top$ are used to account for extra-Poisson
  variation or spatial correlation due to latent or unmeasured
  risk factors. In the following, we will review four models
that are commonly used for~$\mathbf{b}$.

\section{Modelling the spatial dependency structure}\label{sec:spatialModels}

In general, it seems reasonable to assume that areas that are close in space show
more similar disease burden than areas that are not close. To
define what
``close'' means we need to set up a neighbourhood structure.  A common
assumption is to regard areas $i$ and $j$ as neighbours if they
share a common border, denoted here as $i \sim j$.
This seems reasonable if the regions are equally sized and arranged
in a regular pattern \cite{wakefield-2007}. We further denote the
set of neighbours of region $i$  by $\delta i$ and its size by
$n_{\delta i}$.

\subsection{Review of common models}\label{sec:review}
\subsubsection{The Besag model}\label{sec:besag}
One of the  most popular approaches to model spatial correlation is an
intrinsic Gaussian Markov random field (GMRF) model
\cite{besag-etal-1991}, here referred to as the Besag model.
The conditional distribution for $b_i$ is
\begin{gather}
   b_i \mid \mathbf{b}_{-i}, \tau_b \sim
   \mathcal{N}\left(\frac{1}{n_{\delta i}} \sum_{j \in \delta i} b_j,
       \frac{1}{n_{\delta i} \tau_b}\right), \label{eq:besag}
\end{gather}
where $\tau_b$ is a precision parameter and $\mathbf{b}_{-i} = (b_1,
\ldots, b_{i-1}, b_{i+1}, \ldots, b_n)^\top$. The
mean of $b_i$ equals  the mean of the effects
over all neighbours, and the  precision is proportional to the number of
neighbours. The joint distribution for $\mathbf{b}$ is
\begin{align*}
  \pi(\mathbf{b}\mid \tau_b) &\propto \exp \left(-
    \frac{\tau_b}{2} \sum_{i \sim j} (b_i -b_j)^2 \right)
 \propto \exp \left(-
    \frac{\tau_b}{2} \mathbf{b}^\top \mathbf{Q} \mathbf{b} \right),
\end{align*}
where the precision matrix $\mathbf{Q}$ has the entries
\begin{gather}
  Q_{ij} =  \begin{cases}
    n_{\delta i} & i = j,\\
    -1 & i \sim j, \\
    0 & \mbox{else.}
\end{cases}\label{eq:besagQ}
\end{gather}
The model is intrinsic in the sense that $\mathbf{Q}$ is singular,
i.e. it has a non-empty null-space  $\mathbf{V}$, see Rue and Held (Section 3) for
details \cite{gmrfbook}.  For any map, $\mathbf{1}$, denoting a vector of length $n$ with only
ones, is an eigenvector of $\mathbf{Q}$ with eigenvalue $0$. 
Hence the density is invariant to the addition of a constant.  If the spatial map has islands the
definition of the null-space is more complex and depends on how
islands are defined, i.e. whether they form a new unconnected graph or
are linked to the main land, see Section 5.2.1 in Hodges \citep{hodges-2013}.
The rank deficiency of the Besag model is equal to the number of
connected subgraphs. The rank deficiency is equal to one if all
regions are connected. Hence, the density for the Besag model is 
\begin{equation*}
  \pi(\mathbf{b}\mid \tau_b) = K \tau_b^{(n-I)/2}\exp \left(-
    \frac{\tau_b}{2} \mathbf{b}^\top \mathbf{Q} \mathbf{b} \right),
\end{equation*}
where $I$ denotes the number of connected subgraphs and $K$ is a
constant \citep{hodges-2013}.
To prevent confounding with the intercept, sum-to-zero constraints are
imposed on each connected subgraph. 

\subsubsection{The BYM model}

The Besag model only assumes a spatially structured component and
cannot take the limiting form that allows for no spatially structured
variability.
Hence, unstructured random error or pure overdispersion within area $i$, will be modelled as spatial correlation, giving misleading
parameter estimates \cite{breslow-etal-1998}.

To address this issue, the Besag-York-Molli{\'e} (BYM) model  \cite{besag-etal-1991}
decomposes the regional spatial effect $\mathbf{b}$ into a sum of an unstructured
and a structured spatial component, so that $\mathbf{b} = \mathbf{v} + \mathbf{u}$.
Here, $\mathbf{v} \sim \mathcal{N}(\mathbf{0}, \tau_v^{-1}\mathbf{I})$
accounts for pure overdispersion, while  $\mathbf{u} \sim \mathcal{N}(\mathbf{0},
\tau_u^{-1}\mathbf{Q}^{-})$ is the Besag model (Section
\ref{sec:besag}), whereby $\mathbf{Q}^{-}$ denotes the generalised inverse
of $\mathbf{Q}$. The resulting covariance matrix of
$\mathbf{b}$ is
\begin{gather*}
  \text{Var}(\mathbf{b}\mid \tau_u,\tau_v) =\tau_v^{-1}\mathbf{I} + \tau_u^{-1} \mathbf{Q}^{-}.
\end{gather*}

\subsubsection{The Leroux model}

In the BYM model, the  structured and unstructured
components cannot be seen independently from each other, and are thus
not identifiable
\cite{macnab-2010}. An additional challenge, which makes the choice of
hyperpriors more difficult, is that $\tau_v$ and
$\tau_u$ do not represent variability on the same level. While $\tau_v^{-1}$ can be
interpreted as the marginal variance of the unstructured random effects,
$\tau_u^{-1}$ controls the variability of the structured component
$u_i$, conditional on the effects in its neighbouring areas \cite{bernardinelli-etal-1995,
  wakefield-2007}.
Leroux et al. \cite{leroux-etal-2000} proposed an alternative model
formulation to make the compromise between unstructured and structured
variation more explicit. Here, $\mathbf{b}$ is assumed to follow a
normal distribution with mean zero and covariance matrix
\begin{gather}
  \text{Var}(\mathbf{b}\mid \tau_b, \phi) = \tau_b^{-1}\left((1-\phi) \: \mathbf{I} + \phi \mathbf{Q}\right)^{-1}, \label{eq:leroux}
\end{gather}
where $\phi\in [0,1]$ denotes a mixing parameter. The model reduces to pure overdispersion if $\phi = 0$, and to the Besag
model when $\phi =1$ \cite{lee-durban,macnab-2010}. The conditional expectation of $b_i$, given all other
random effects, results as a weighted average of the zero-mean unstructured model and the mean value of the Besag model \eqref{eq:besag}.
The conditional variance is the weighted average of $1/\tau_b$ and
$1/(\tau_b \cdot n_{\delta i})$.

\subsubsection{The Dean model}

Dean et al. \cite{dean-etal-2001} proposed a reparameterisation of the
BYM model where
\begin{gather}
  \mathbf{b} = \frac{1}{\sqrt{\tau_b}} \left(\sqrt{1-\phi} \:\mathbf{v} +
    \sqrt{\phi} \:\mathbf{u}\right) \label{eq:dean},
\end{gather}
having covariance matrix
\begin{gather}
  \text{Var}(\mathbf{b}\mid \tau_b, \phi) = \tau_b^{-1}\left((1-\phi) \: \mathbf{I} + \phi \mathbf{Q}^{-}\right)\label{eq:varbym2}.
\end{gather}
Equation \eqref{eq:dean} is a reparameterisation of the original BYM
model,
where $\tau_u^{-1} =\tau_b^{-1} \phi$ and $\tau_v^{-1} = \tau_b^{-1}
 (1-\phi)$ \cite{macnab-2010}.
The additive decomposition of the variance is then on the log
relative risk scale. This is in contrast to the Leroux model  \eqref{eq:leroux}, where
the precision matrix of $\mathbf{b}$ resulted as a weighted average of the
precision matrices of the unstructured and structured spatial
components. As a consequence, the additive
decomposition of variance in the Leroux model happens on the log relative risk scale,
conditional on $b_j$, $j \in \delta i$ \cite{leroux-etal-2000}.

\subsection{Why do we need scaling?}\label{sec:scaling}

One issue inherent to all the models described in
Section~\ref{sec:review}, is that the spatially structured component is
not scaled. However, as discussed in S{\o}rbye and
Rue\cite{sorbye-rue-2014}, scaling is crucial to facilitate hyperprior
assignment and guarantee that hyperpriors used in
one application have the same interpretation in another application.

The Besag model is an intrinsic GMRF  and penalises local deviation from its null
space, which is a constant level in the case of one connected component \cite{gmrfbook}. 
The
hyperprior will control this local deviation and, as such,  influence
the smoothness of the estimated spatial effects. If the estimated field is too
smooth, the precision is large and potential spatial variation might be
blurred. On the other hand,  if the precision is too small the model might overfit due to large local variability \citep{sorbye-rue-2014}.

In the following we consider only the structured effect $\mathbf{b}$
as introduced in the Besag model in Section \ref{sec:besag}.  In general, the marginal variances
$\tau_b^{-1}[\mathbf{Q}^{-}]_{ii}$ depend on the graph
structure reflected by the structure matrix $\mathbf{Q}$ \cite{sorbye-rue-2014}.
This can be illustrated by calculating a generalized variance, computed as
the geometric mean (or some other
typical value) of the marginal variances
 \begin{eqnarray}
  \sigma_{\text{GV}}^2(\mathbf{u}) &= &  \exp \left(\frac{1}{n}\sum_{i=1}^n
    \log\left(\frac{1}{\tau_b} [\mathbf{Q}^{-}]_{ii}\right) \right)
    = \frac{1}{\tau_b} \exp \left(\frac{1}{n}\sum_{i=1}^n
    \log([\mathbf{Q}^{-}]_{ii}) \right). \label{eq:cv}
\end{eqnarray}
The generalized variance  for two different graph structures are typically not equal, even when the graphs have the same number of regions.

\begin{figure}
\centering
\includegraphics[width=0.5\textwidth]{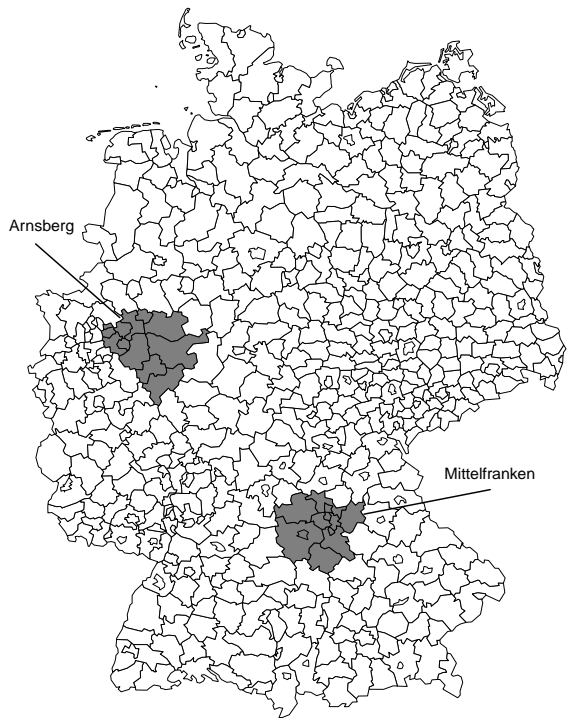}
\caption{Map of Germany separated in $544$ districts. Marked in gray
  are the administrative districts Arnsberg in the federal state
  Nordrhein-Westfalen  and Mittelfranken  in Bayern. Both areas
  consist of 12 districts.\label{fig:germany} }
\end{figure}

As an example, Figure~\ref{fig:germany} shows
two administrative regions of Germany, Mittelfranken and  Arnsberg,
that are both fully connected. Assume that we fit models to these two regions separately, including
a spatially structured effect in the models. As both regions have 12
districts, we may be tempted to use identical  priors for the precision
$\tau_b$ in these two cases.  However, the prior will penalise the local deviation from
a constant level, differently. Specifically, if $\tau_b = 1$ in \eqref{eq:cv}, it follows that
$\sigma_{\text{GV}}^2(\mathbf{b}) \approx 0.29$ for Mittelfranken and
$\sigma_{\text{GV}}^2(\mathbf{b}) \approx 0.40$ for Arnsberg.
If we fit a similar model to all the $544$ districts in Germany, we obtain
$\sigma_{\text{GV}}^2(\mathbf{b}) \approx 0.56$.
These  differences, reflecting a characteristic level for the marginal
variances, might be even more striking for other applications. We find
that the value of the generalised variance is larger for more complex
neighbourhood structures, for example for the map of the 3111 counties
in the continental United States we obtain
$\sigma_{\text{GV}}^2(\mathbf{b}) \approx 4.78$.
 In order to
unify the interpretation of a chosen prior for $\tau_b$ and make it  transferable
between applications, the structured effect needs to be scaled so that
$\sigma_{\text{GV}}^2(\mathbf{b}) = 1/\tau_b$. This implies that $\tau_b$
represents the precision of the (marginal) deviation from a constant
level, independently of the underlying graph.

This issue of scaling applies to all IGMRFs \cite{sorbye-rue-2014}. In practice, the scaling might be costly, as it involves inversion of a singular
matrix. The generalised inverse can be used, but the relative tolerance
threshold to detect zero singular values will be crucial. Knowing the
rank deficiency $r$ of the matrix, the $r$ smallest eigenvalues could
be removed. IGMRFs have much in common with GMRFs conditional on
linear constraints, see~\citep[Section 3.2]{gmrfbook}. Hence, if we know
the null space $\mathbf{V}$ of $\mathbf{Q}$,  we can use
sparse matrix algebra to  compute $[\mathbf{Q}^{-}]_{ii}$.
 The marginal variances can be computed based on
$\mathbf{b} \mid \mathbf{V}^\top \mathbf{b} = \mathbf{0}$. First, $\mathbf{Q}$ is
made regular by adding a small term $\epsilon$ to the diagonal and then
 $\text{Diag}(\text{Var}(\mathbf{b} \mid \mathbf{V}^\top \mathbf{b}=\mathbf{0}))$ is computed according to \citep[Section 3.2, Equation 3.17]{gmrfbook}. In this way,  we take advantage of the sparse matrix structure of
   $\mathbf{Q}$. Extracting the variance components, we compute
   $\sigma_{\text{GV}}^2$ and use this as a factor to scale
   $\mathbf{Q}$. 
The R-package {\tt INLA}, available from \url{www.r-inla.org}, offers
a function {\tt inla.scale.model} which takes as argument 
any singular precision matrix $\mathbf{Q}$ and  the linear constraints
spanning the null-space of $\mathbf{Q}$. The scaled precision matrix,
where the geometric mean of the marginal variances is equal to one, is
returned. 
For the  Besag model on a connected graph we have
$\mathbf{V} =   \mathbf{1}$, 
so that we can scale the matrix in {\texttt R} using
\begin{Sinput}
R> Q = inla.scale.model(Q,
     constr=list(A=matrix(1, nrow=1, ncol=n), e=0))
\end{Sinput}

We would like to note that scaling is also crucial if the goal is
hypothesis testing involving an IGMRF, as for example in Dean et al.~\cite{dean-etal-2001}, as otherwise the
critical region would be naturally dependent on the scale of the
random effect. This also applies outside a Bayesian setting.

\subsection{A modified BYM model accounting for scaling}

Simpson et al. \cite{simpson-etal-2015} propose a modification of the Dean model which
addresses both  the identifiability and scaling issue of the BYM model. The model uses a scaled structured
component $\mathbf{u}_\star$, where $\mathbf{Q_\star}$ is the
precision matrix of the Besag model, scaled according to Section \ref{sec:scaling}. This gives a modified version of the random effect
\begin{gather}
  \mathbf{b} = \frac{1}{\sqrt{\tau_b}} \left(\sqrt{1-\phi} \:\mathbf{v} +
    \sqrt{\phi} \:\mathbf{u}_\star\right) \label{eq:bym2},
\end{gather}
having covariance matrix
\begin{gather}
  \text{Var}(\mathbf{b}\mid \tau_b, \phi) = \tau_b^{-1}\left((1-\phi) \: \mathbf{I} + \phi \mathbf{Q}_\star^{-}\right)\label{eq:varbym2}.
\end{gather}

Breslow et al. \cite{breslow-etal-1998} criticised the BYM model for
obscuring the conditional mean and conditional variance. Using the new
parameterisation,  $ \text{Var}(\mathbf{b}\mid \tau_b, \phi)$ has a clear and intuitive interpretation, and interpretations in terms of the conditional distribution of $b_i$ given $b_j$, $j \in \delta i$, are avoided.

Similarly to the Leroux and the Dean model, \eqref{eq:bym2}
emphasises
the compromise between pure
overdispersion and spatially structured correlation, where $0 \leq
\phi \leq 1$ measures the proportion of the marginal variance
explained by the structured effect. The model
reduces to pure overdispersion for $\phi = 0$  and to the Besag model,
i.e.~only spatially structured risk, when $\phi = 1$. 
More importantly, using the standardised $\mathbf{Q}_\star^{-}$ the marginal variances 
will be approximately equal to  $(1-\phi)/\tau_b + \phi/\tau_b$. Approximation
holds since $\mathbf{Q}^-$ by construction does not  give the
same marginal variances for all regions, so that the standardisation
only holds on the overall level of the generalised variance $\sigma^2_{\text{GV}}$. With the 
scaled structured effect $\mathbf{u}^\star$, the random effect
$\mathbf{b}$ is also scaled and the prior imposed on $\tau_b$ has the
same interpretation, being transferred between graphs. Furthermore,
the hyperparameters $\tau_b$ and $\phi$ are now interpretable and no
longer confounded. It should be noted that in contrast to the Dean
model, the Leroux model cannot be scaled since by construction the
scaling would depend on the value of $\phi$.

\subsection{Parameterisation preserving sparsity}

A challenge of the new model formulation in
\eqref{eq:bym2} and \eqref{eq:varbym2} is that the covariance matrix
involves the generalised inverse $\mathbf{Q}_\star^{-}$. The
precision matrix, defined as the inverse of \eqref{eq:varbym2}, is then no
longer sparse. Sparsity properties are important to gain computational efficiency,  both implementing the model in the
INLA framework \cite{rue-etal-2009},  or using block updating in Markov chain Monte Carlo methods. To keep the sparsity of the precision matrix,  we propose an augmented
parameterisation.

Recall that $\mathbf{u}_\star \sim
\mathcal{N}(\mathbf{0}, \mathbf{Q}_\star^{-})$, where $\mathbf{Q}_\star$
is scaled, and $\mathbf{v} \sim \mathcal{N}(\mathbf{0}, \mathbf{I})$ in  \eqref{eq:bym2}. Let
$\mathbf{w} = \begin{pmatrix}
  \mathbf{w}_1^\top &
  \mathbf{w}_2^\top
\end{pmatrix}^\top$, where $\mathbf{w}_2 = \mathbf{u}_\star$ and
$\mathbf{w}_1 = \mathbf{b}$.  Thus,
\begin{gather*}
 \mathbf{w}_1 \mid \mathbf{w}_2  \sim
 \mathcal{N}\left(\sqrt{\frac{\phi}{\tau_b}} \mathbf{w}_2,
   \frac{1-\phi}{\tau_b} \mathbf{I}\right).
\end{gather*}
As $\pi(\mathbf{w}) = \pi(\mathbf{w}_1 \mid \mathbf{w}_2) \pi(
\mathbf{w}_2)$, it follows that $\mathbf{w}$ is normally distributed
with mean $\mathbf{0}$ and precision matrix
\begin{gather*}
\begin{pmatrix}
 \frac{\tau_b}{1-\phi} \mathbf{I} & - \frac{\sqrt{\phi
     \tau_b}}{1-\phi} \mathbf{I}\\
- \frac{\sqrt{\phi
     \tau_b}}{1-\phi} \mathbf{I} & \mathbf{Q}_\star + \frac{\phi}{1-\phi}\mathbf{I}
\end{pmatrix}.
\end{gather*}
The marginal of $\mathbf{w}_1$ then has the correct distribution.
Working with this parameterisation,  sparsity is warranted and the
structured component is given directly by $\mathbf{w}_2$, i.e. the second half of the
vector $\mathbf{w}$.

\section{Hyperprior choice: Penalised-complexity priors}\label{sec:pcprior}

In applying the original BYM model formulation, priors are usually
specified separately for $\tau_u$ and $\tau_v$ and the prior choices are
frequently rather ad-hoc \cite{bernardinelli-etal-1995}. A reasonable alternative,
is to express prior beliefs in terms of the total variability of the model
\cite{wakefield-2007}, which also facilitates clear interpretation of the hyperparameters.

Here, we follow Simpson et al. \cite{simpson-etal-2015} and apply the 
framework of penalised complexity (PC) priors, which
allows us to distribute the total variance to the components of the modified
BYM model. The construction of PC priors is  based on four
principles: 1. {\em Occam's razor} --- simpler models should be
preferred until there is enough support for a more complex
model. 2. {\em Measure of complexity}  --- The Kullback-Leibler
distance (KLD) is used to measure increased complexity. 3. {\em
  Constant-rate penalisation} --- The deviation from the simpler model
is penalised using a constant decay rate. 4. {\em User-defined
  scaling}  --- The user has an idea of a sensible size of the parameter (or
any transformed variation) of interest. 
For ease of readability, we provide the main ideas and results using
the  PC framework, and refer to Simpson et
al. \cite{simpson-etal-2015} Section~6, Appendices~A.2 and A.3 for further mathematical details.

Hyperpriors are formulated in two levels. First, a prior for
$\tau_b$ is defined, which will control the marginal variance contribution of the weighted sum
of $\mathbf{v}$ and $\mathbf{u}_\star$. Second, the marginal variance is distributed to the components  $\mathbf{v}$ and
$\mathbf{u}_\star$ by defining a suitable prior for the mixing
parameter $\phi$. Note that testing for the risk decomposition,
as proposed in MacNab \cite{macnab-2010}, is implicated using the
PC-prior framework at this level. The PC prior will
automatically shrink towards $\phi=0$, which refers to no spatial
smoothing. The  modified BYM
model is then seen as a flexible extension of two dependent base models, which it will shrink to if not differently
supported by the data.

\subsection{Controlling the spatial variance contribution} \label{sec:pc-prec}
The simplest model, referred to as a first-level base model, is to have a constant
relative risk over all areas, i.e. no spatial variation or
equivalently infinite precision.
All $n$ regions have the same disease burden.
Following the principles of the PC prior framework, we penalise
model complexity in terms of a measure of information-theoretic deviation from the flexible model to the base model,  which is a function of the  Kullback-Leibler divergence
\begin{gather}
\text{KLD}(\tau_b)= \text{KLD}(\mathcal{N}(0,
\mathbf{\Sigma}_{\text{flex}}(\tau_b))  \mid \mathcal{N}(0,
\mathbf{\Sigma}_{\text{base}})). \label{eq:KLD}
\end{gather}
KLD  measures the information lost using the
base model to approximate the more flexible model. To facilitate interpretation, this divergence is transformed to a unidirectional measure of distance defined by 
\begin{equation*}
d(\tau_b) = \sqrt{2\text{KLD}(\tau_b)}.
\end{equation*}
In our case, $\mathbf{\Sigma}_{\text{flex}}(\tau_b)=
\text{Var}(\mathbf{b}\mid \tau_b, \phi)$ as defined in (\ref{eq:varbym2}),
while the covariance matrix of the base model is
$\mathbf{\Sigma}_{\text{base}}=\mathbf{0}$, reflecting infinite
precision. Details regarding the computation of \eqref{eq:KLD} are
presented in Simpson et al. \cite{simpson-etal-2015} and for completeness
provided in~\ref{app:pc}.  On the distance scale the meaning of a prior is much more
clear.  The prior should have mode at zero, i.e. at the base model, and decreasing density
with increasing distance. There are different priors that fulfill
these properties, but differ in the way they decrease with increasing
distance. Since we are not in a variable selection setting, we follow
the recommendation of Simpson et al. \cite{simpson-etal-2015} and use
the exponential 
distribution, which supports constant-rate penalisation.  Applying the ordinary change
of variable transformation, a type-2 Gumbel distribution\citep{simpson-etal-2015}
\begin{gather*}
  \pi(\tau_b) = \frac{\theta}{2} \tau_b^{-3/2} \exp\left(-\theta
    \tau_b^{-1/2}\right)
\end{gather*}
is obtained as prior for $\tau_b$.

\begin{figure}
\centering
\includegraphics[width=0.9\textwidth]{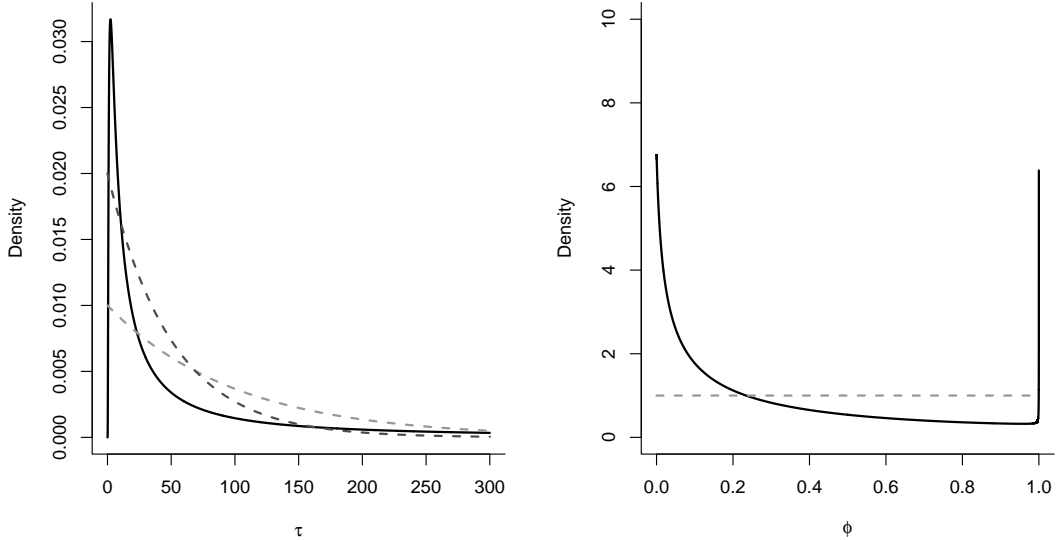}
\caption{Left panel: PC prior for $\tau_b$ (solid) with parameters
  $U=1$ and $\alpha = 0.01$, and two gamma priors where for both the
  shape parameter is equal to one and the rate parameter is either
  0.01 (dashed, light-gray) or 0.02 (dashed, dark-gray). Right panel:
   PC prior for the mixing parameter $\phi$ with $U = 0.5$ and $\alpha =
  2/3$ (solid), and a uniform prior on (0,1) (dashed). }\label{fig:prior}
\end{figure}

The given prior corresponds to an exponential prior with rate $\theta$
for the standard deviation $1/\sqrt{\tau_b}$. Hence, to infer $\theta$
we can use the probability statement $\text{Prob}((1/\sqrt{\tau_b}) >
U) = \alpha$, which gives $\theta = - \log(\alpha)/U$. It is not
obvious how to interpret $1/\sqrt{\tau_b}$ epidemiologically. However,
in the BYM2 model, $\tau_b$ represents the marginal precision, and as
such, can be related to the total residual relative risk for which an
intuitive interpretation is available \cite{wakefield-2007}. For
example, an assumed probability of $0.99$ of having residual relative
risks  smaller than 2, corresponds approximately to
$\text{Prob}((1/\sqrt{\tau_b}) > 1) = 0.01$. Figure~\ref{fig:prior}
(left panel) shows the resulting prior compared to two gamma
distributions,  which will be used in the simulation study in
Section~\ref{sec:application}. Please note that the use of gamma priors is questionable as the base model, i.e.
distance equal to zero, has no density mass and can therefore never be
obtained. We refer to Simpson et al.\cite{simpson-etal-2015} for further discussions about overfitting of priors.

\subsection{Distributing the spatial variance  contribution}
Assume now that the relative risk has spatial variation. This implies that 
the disease burden over the regions is not constant. Conditioned on a fixed marginal variance, the
second-level base model is defined as having only unstructured spatial noise and no spatial
dependency, i.e. $\phi=0$. By increasing the value of $\phi$, spatial dependency is gradually blended into the model and the model component explaining most of the variance shifts from
$\mathbf{v}$ to $\mathbf{u}_\star$\cite{simpson-etal-2015}.

To derive the PC prior for $\phi$, we first compute the KLD of the base model from
the flexible model, see~Appendix~\ref{app-mixing}. The covariance matrices for the base and flexible models are
$\mathbf{\Sigma}_{\text{base}} =  \mathbf{I}$ and
$\mathbf{\Sigma}_{\text{flex}}(\phi) =  (1-\phi) \: \mathbf{I} +
\phi\mathbf{Q}_\star^{-}$, respectively. 
Analogously as for $\tau_b$, we assign an exponential prior with parameter $\lambda$ to
the distance scale $d(\phi) = \sqrt{2\mbox{KLD}(\phi)}$. For simplicity, this prior is
transformed to give a prior on $\text{logit}(\phi)$, as $\phi$ is
defined on a bounded interval. Reasonable values of  $\lambda$ can be
inferred in a similar way as in Section~\ref{sec:pc-prec}, using a
probability statement $\text{Prob}(\phi < U) = \alpha$. In this way
it is straightforward to specify both strong or weak prior
information. 
The parameter $\phi$ has a more direct  interpretation  than
the precision parameter $\tau_b$, representing a fraction of the total variance which can be attributed to spatial dependency structure. A reasonable formulation might
be  $\text{Prob}(\phi < 0.5) = 2/3$, which gives more density mass to
values of $\phi$ smaller than $0.5$. This can be seen as a
conservative choice, assuming that the unstructured random effect
accounts for more of the variability than the spatially  structured
effect.

In contrast to the precision parameter $\tau_b$, the resulting PC prior for $\phi$
is not available in closed form.
Within INLA, the function {\tt INLA:::inla.pc.bym.phi} can be used
to compute the prior for $\phi$ (on logit-scale) for a specific
Besag neighbourhood matrix~$\mathbf{Q}$.
The scaling of the graph is implemented by default.
Thus, the prior can be obtained in tabulated form and also
incorporated in other software packages than INLA.
Figure~\ref{fig:prior} (right panel)  shows the prior for $\phi$ with  $U = 0.5$ and $\alpha=2/3$,
generated in {\tt R} using
\begin{Sinput}
R> log.prior = INLA:::inla.pc.bym.phi(Q = Q, rankdef = 1,
R>    u = 0.5, alpha = 2/3)
R> phis = 1/(1+exp(-seq(-10, 10, len=10000)))
R> plot(phis, exp(log.prior(phis)), type="l",
R>     lwd = 2,  bty = "l", ylim = c(0,10),
R>     xlab = expression(phi), ylab = "Density")
\end{Sinput}
For comparison, the figure also displays the commonly used uniform prior for $\phi$.

\section{Applications}\label{sec:application}

\subsection{Does the modified BYM model shrink to the respective base models?}

Various types of cancer data, observed in the $544$ districts of
Germany (Figure~\ref{fig:germany}), have been widely analysed in
the literature, both to study the epidemiology
\cite{natario-knorrHeld-2003} and to test
methodological developments \cite{knorr-rue-2002}. Simpson et
al. \cite{simpson-etal-2015} re-analyse a larynx cancer dataset, assigning the PC priors shown in
Figure~\ref{fig:prior} to the hyperparameters. The modified BYM model
(\ref{eq:bym2}), here referred to as the BYM2 model, is clearly seen to
learn from the data, leading to a posterior marginal concentrated
around 1 for~$\phi$. 

\subsubsection{Results}\label{sec:results}
Here, we perform a simulation study to investigate whether the BYM2
model shrinks towards the two different base models described in
Section~\ref{sec:pcprior}. Furthermore, we assess the behaviour under 
a perfectly structured risk surface. The simulations are based on the
neighbourhood structure of $366$ districts of Sardinia, for which
a real case study will be analysed in Section~\ref{sec:sardinia}. 
Two hundred datasets were simulated under
nine different scenarios, using three different types of risk surfaces
and three different disease prevalences. The model we simulate
from assumes that the log relative risk is given as
$\log(\theta_i) = \mu + \frac{1}{\sqrt{\tau_b}} b_i$, $i=1, \ldots, 366$, where $\mathbf{b}$
is a zero-mean normal distributed random effect controlled by the
precision parameter $\tau_b$. In the simulations, we assume either 
a constant risk over all regions, i.e. $1/\sqrt{\tau_b}=0$,  an
independent area-specific risk setting with $1/\sqrt{\tau_b}=0.5$ and
$\mathbf{b} \sim \mathcal{N}(\mathbf{0}, \mathbf{I})$,  or a spatially
structured area-specific risk setting with $1/\sqrt{\tau_b}=0.5$ and
$\mathbf{b} \sim \mathcal{N}(\mathbf{0}, \mathbf{Q}_\star^{-})$.
  We alter the expected number of cases $E_{i}$ to analyse whether
  the model performance changes when applied to diseases with
  different underlying disease prevalence. We set all $E_{i}$ equal
  to either $15$, $60$ or $200$ and analyse all datasets using the
  {\tt INLA} package in~{\tt R}.

Table~\ref{tab:simBYM2} shows the posterior mean estimates for the
intercept and the two hyperparameters, also including the average standard deviations of the
estimates in parentheses. 
\begin{table}
\small\sf\centering
\caption{Mean values of the intercept and hyperparameters,
 for the BYM2 model for varying disease
  prevalence, using either a constant, spatially-varying unstructured
  risk surfaces or spatially-varying structured risk surfaces. PC
  priors are used for the weighting parameter $\phi$ and for the
  precision parameter $\tau_b$. Standard deviations are
  provided in parentheses.  Results are based on $200$ simulations
for each setting, where the true  parameter values are provided in the table.} \label{tab:simBYM2}
\begin{tabular}{lccc}
  \toprule
  & $\mu$  & $1/\sqrt{\tau}$ & $\phi$ \\
  \midrule
  \multicolumn{4}{l}{Constant risk ($1/\sqrt{\tau_b} = 0$, $\phi=0$):} \\[.2cm]
E = 15 & -3.78E-04 (0.01) & 0.02 (0.02) & 0.26 (0.05) \\ 
  E = 60 & -9.15E-04 (0.01) & 0.01 (0.01) & 0.26 (0.05) \\ 
  E = 200 & -4.36E-04 (0.004) & 0.01 (0.01) & 0.26 (0.04) \\ 
  \multicolumn{4}{l}{Area-specific independent risk ($1/\sqrt{\tau_b}
    = 0.5$, $\phi = 0$):}  \\[.2cm]
E = 15 & 1.62E-03 (0.03) & 0.50 (0.02) & 0.05 (0.03) \\ 
  E = 60 & -1.94E-03 (0.03) & 0.51 (0.03) & 0.04 (0.02) \\ 
  E = 200 & 1.21E-03 (0.03) & 0.50 (0.02) & 0.04 (0.02) \\ 
  \multicolumn{4}{l}{Area-specific correlated risk ($1/\sqrt{\tau_b}
    = 0.5$, $\phi = 1$):}  \\[.2cm]
E = 15 & 2.14E-03 (0.01) & 0.47 (0.04) & 0.90 (0.06) \\ 
  E = 60 & 9.11E-06 (0.01) & 0.49 (0.05) & 0.93 (0.04) \\ 
  E = 200 & 4.70E-05 (0.004) & 0.49 (0.07) & 0.93 (0.04) \\ 
   \bottomrule
 \end{tabular}
\end{table}
The constant risk surface scenario corresponds to the first-level base
model described in Section~\ref{sec:pcprior}. The resulting estimates are reasonable for all parameters. Specifically, the standard
deviation $1/\sqrt{\tau_b}$ is close to zero, and decreases with higher prevalence.
As these datasets give no information about the mixing parameter, the prior will dominate the posterior estimates of $\phi$.
The assumption of independent, area-specific spatial variation of the risk surface, corresponds to the second-level base model in Section~\ref{sec:pcprior}.  Again, the posterior mean parameter estimates are seen to be close to the true values. In particular, the model is seen to shrink towards the base model as the estimated mixing
parameters are very close to zero, distributing all of the spatial variation to the
unstructured component. In the case of a spatially structure surface
risk,  the intercept and $1/\sqrt{\tau_b}$ are well estimated.
The mixing parameter $\phi$ is estimated slightly lower when
$E_{i}=15$ compared to $E_{i}=60$ and $E_{i}=200$, where the estimates
are closer to $1$. This illustrates the learning ability of the model
to support a more complex model if indicated by the data. 

We repeated the simulation study using a uniform prior for the mixing
parameter $\phi$. The obtained results are comparable, see
Table~\ref{tab:simBYM2_unif} in~\ref{app:figtab}. Differences
occur in the case of a simulated constant risk surface where the estimates
of the mixing parameter are  now close to
$0.5$, as the estimate is dominated by the new prior. 
In the case of a spatially unstructured or structured risk, the
obtained  parameter estimates are seen to be robust to the
prior choice for the mixing parameter.

\subsubsection{Comparison to other spatial latent models}
The main benefit of the BYM2 model is that it allows for
an intuitive parameter interpretation and facilitates prior
assignment. Performance in terms of model choice criteria is regarded
secondary and will be assessed in this section. To compare the BYM2
model with existing approaches, we fit the spatial
latent models presented in Section~\ref{sec:review} to the simulated data. We
also include a model having only an unstructured spatial component,
referred to as the iid-model. For the BYM2 model we use either a PC
prior with $U=0.5$ and $\alpha=2/3$, denoted as BYM2 (PC), or a
uniform prior, denoted as BYM2 (unif), for the mixing parameter.

Table~\ref{tab:simAll} illustrates the resulting posterior parameter estimates,
assuming a constant risk surface and $E=60$. The table also reports three model choice criteria for this scenario: The
root-mean squared error (RMSE), computed as the average of
$\sqrt{\tfrac{1}{366} \sum_{i=1}^{366}(y_i/E_i - \hat{\theta}_i)^2}$ over all 200
simulations with $\hat{\theta}_i$ denoting the posterior mean fitted risk, the average deviance information criterion (DIC)
\cite{spiegelhalter-etal-2002}, and the average proper logarithmic scoring rule
(LS) \cite{gneiting-raftery-2007}. For all three criteria, lower
values imply better model properties. Note that the parameter estimates of
the BYM model are not included here as it is not parameterised in
terms of one precision and one mixing parameter. However, we will
consider it in terms of model choice, see Figure~\ref{fig:cpo} -- \ref{fig:dic}.

The estimated parameter values in Table~\ref{tab:simAll} have to be
interpreted with great care. 
Even though the different models look comparable and the parameters have the same name, their interpretations are not equivalent. A first issue is that
we use different priors for the hyperparameters. Specifically, we assume that  $\phi \sim
\text{Unif}(0,1)$ for the Leroux and Dean model, as this represents a popular
non-informative prior choice in the literature
\citep{wakefield-2007,macnab-2010,ugarte-etal-2014}. Further, we assume a 
$\Gamma(\text{shape}=1, \text{rate}=0.01)$ distribution for the precision
parameter in the iid-model. For the spatially structured term, we took the graph structure of
Sardinia into account and use a $\Gamma(\text{shape}=1, \text{rate}=0.02)$
\cite{bernardinelli-etal-1995} prior. However, also the different
parameterisation of the Leroux and Dean model imply that the parameters
are not comparable.  It is only when $\phi=0$ or $\phi =1$, these models
reduce to the same model (assuming the same prior for the
precision parameter).

The second issue is that none of the models are properly scaled, except the BYM2 model.  Specifically, the 
precision parameter in the Leroux and Dean model does not represent
the marginal precision.  It is still confounded with $\phi$ and hence not
comparable to the BYM2 model. The only parameter estimates that are
comparable are those from the iid-model and the BYM2 model. The Besag
model can be scaled, and so can the  Dean model leading to
the parameterisation of the BYM2 model. However, the Leroux model cannot be
scaled as any scaling would depend on the mixing parameter
$\phi$. Hence,  the precision parameter cannot be interpreted as
marginal precision as it depends on the
underlying graph structure and is in addition confounded with the mixing
parameter. 

Even though the parameters of the models have different interpretations and priors are chosen differently, the
model choice criteria give quite similar average values, for all the
models and only the DIC values seem to slightly favour the two BYM2
models. This means that the BYM2 performs at least as well as the other
models and changing the prior for the mixing parameters leads to the
same results. This is also confirmed inspecting the
logarithmic scores and  DIC values over all simulations, in terms of boxplots, see
Figures~\ref{fig:cpo} and \ref{fig:dic} in~\ref{app:figtab}, respectively. Results for these
two model choice criteria coincide well and indicate that the
Besag model performs worse than the other models when simulating an
independent area-specific risk surface, where the iid-model is
naturally but only marginally favoured. In contrast, the iid model performs
worse when simulating a spatially-structured surface, where the Besag
model is marginally favoured for low values of $E_{i}$. The other
models perform almost identically in both of these simulation
settings. Simulating a constant-risk surface, all models behave similarly
for $E_{i}=15$. For increasing numbers of expected cases, the BYM2
model seems to be slightly beneficial. The figures only include
the boxplots obtained by the BYM2 model using a PC prior for $\phi$,
as results using a uniform prior are visually identical.

\begin{table}
\small\sf\centering
\caption{Mean values of the intercept and hyperparameters,
  root mean square error (RMSE) for the fitted risk, DIC and logarithmic score (LS) when using
  different spatial latent models. Standard deviations are
  provided in parentheses.  Results are based on $200$ simulations
where the true values are $\mu = 0$, $1/\sqrt{\tau} = 0$, $\phi = 0$,
$E=60$. Be aware that parameter estimates might not be directly
comparable between models.} \label{tab:simAll}
\begin{tabular}{lccccccc}
  \toprule
  & $\mu$  & $1/\sqrt{\tau}$ & $\phi$ & RMSE & DIC & LS  \\
  \midrule
  iid-model & -1.65E-03 (0.01) & 0.05 (0.004) & 0.00 (-) & 0.11 & 2545 & 3.47 \\ 
  Besag & -1.26E-03 (0.01) & 0.07 (0.004) & 1.00 (-) & 0.12 & 2545 & 3.48 \\ 
  Leroux & -1.51E-03 (0.01) & 0.07 (0.01) & 0.56 (0.08) & 0.11 & 2545 & 3.48 \\ 
  Dean & -1.69E-03 (0.01) & 0.06 (0.004) & 0.60 (0.12) & 0.11 & 2547 & 3.48 \\ 
  BYM2 (unif) & -8.39E-04 (0.01) & 0.01 (0.01) & 0.46 (0.05) & 0.12 & 2540 & 3.47 \\ 
  BYM2 (PC)  & -9.15E-04 (0.01) & 0.01 (0.01) & 0.26 (0.05) & 0.12 & 2540 & 3.47 \\ 
   \bottomrule
\end{tabular}
\end{table}

\subsection{Analysis of insulin-dependent diabetes mellitus in Sardinia}\label{sec:sardinia}

We re-analyse insulin-dependent diabetes mellitus (IDDM) counts in
Sardinia, see Bernardinelli et al.~\citep{bernadinelli-etal-1997} and
Knorr-Held and Rue \cite{knorr-rue-2002} for previous analyses. This dataset
has low counts. Only $619$ cases were registered for $n=366$ districts
in the period $1989$--$1992$, for the population aged $0$--$29$ years
and there are no covariates. Spatial smoothing is essential to reduce
the large variance in the estimates, and to get a realistic picture of
the underlying risk surface. Held \cite{held-2004} has demonstrated that a
constant risk is well supported by the posterior distribution, 
and based on this we would expect shrinkage towards the first-level base model.

\begin{figure}
\centering
\includegraphics[width=.7\textwidth]{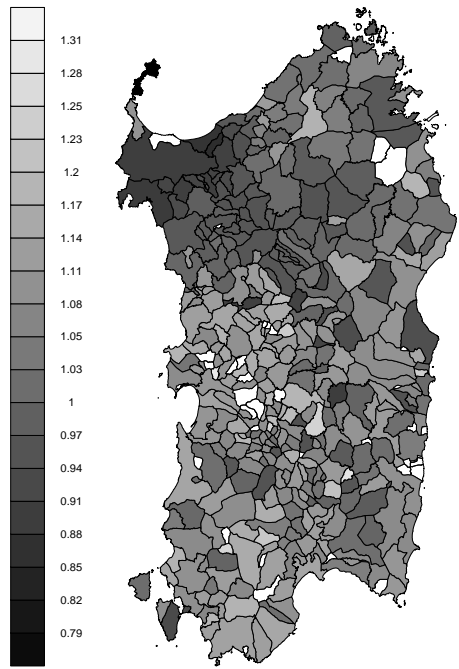}
\caption{Posterior mean of the relative risk for IDDM counts in
  Sardinia using the BYM2 model.}\label{fig:sardiniaRR}
\end{figure}

We use $U = 0.5$ and $\alpha=2/3$ for the prior of $\phi$. For
$\tau_b$ we use $U=0.2/0.31$ and $\alpha=0.1$ which corresponds to a
marginal standard deviation of $0.2$. The R-code to fit the BYM2 model
using the INLA-package is explained in~\ref{app:rcode}.
The posterior estimates are shown in Table~\ref{tab:sardiniaEst}
(first quarter). We
see that the precision estimate is rather high supporting a constant log
relative risk as a zero log relative risk is well within all the
posterior credible intervals for $\{\eta_i\}$. The
posterior marginal is mostly uniform giving essentially no information on which
component is describing the variance. Figure~\ref{fig:sardiniaRR}
shows the posterior mean for the relative risk over all $366$ regions.

\begin{table}
\small\sf\centering
\caption{Posterior quantities for the intercept and hyperparameters
  under three different PC priors or a uniform prior for the mixing parameter $\phi$ for IDDM
  counts in Sardinia.}\label{tab:sardiniaEst}
\begin{tabular}{rrrrrrr}
  \toprule
 & Mean & SD & $2.5\%$ quant. & Median & $97.5$ quant. & Mode \\ 
  \midrule  
&\multicolumn{4}{l}{Prior for $\phi$: $U=0.5$, $\alpha=2/3$} & \\[.2cm]
$\mu$ & 0.01 & 0.05 & -0.09 & 0.01 & 0.10 & 0.01 \\ 
 $\tau_b$ & 262.79 & 751.60 & 10.14 & 45.36 & 2549.91 & 17.40 \\ 
$\phi$ & 0.48 & 0.31 & 0.01 & 0.48 & 0.98 & 1.00 \\[.2cm]
   &\multicolumn{4}{l}{Prior for $\phi$: $U=0.5$, $\alpha=0.5$} & \\[.2cm] 
 $\mu$ & 0.01 & 0.05 & -0.09 & 0.01 & 0.10 & 0.01 \\ 
 $\tau_b$ & 233.51 & 686.28 & 10.00 & 42.78 & 2241.48 & 17.37 \\ 
$\phi$ & 0.54 & 0.31 & 0.02 & 0.57 & 0.99 & 1.00 \\[.2cm]
   &\multicolumn{4}{l}{Prior for $\phi$: $U=0.5$, $\alpha=0.1$} & \\[.2cm]  
 $\mu$ & 0.01 & 0.05 & -0.08 & 0.01 & 0.11 & 0.01 \\ 
 $\tau_b$ & 201.56 & 616.57 & 9.83 & 40.11 & 1880.89 & 17.27 \\ 
$\phi$ & 0.63 & 0.29 & 0.03 & 0.69 & 1.00 & 1.00 \\[.2cm]
   &\multicolumn{4}{l}{Prior for $\phi$: Unif$(0,1)$} & \\[.2cm]  
 $\mu$ & 0.01 & 0.05 & -0.08 & 0.01 & 0.11 & 0.01 \\ 
 $\tau_b$ & 189.46 & 591.67 & 9.77 & 39.07 & 1734.34 & 17.21 \\ 
$\phi$ & 0.64 & 0.26 & 0.08 & 0.69 & 0.99 & 1.00 \\ 
   \bottomrule
\end{tabular}
\end{table}

\begin{figure}
\centering
\includegraphics[width=.95\textwidth]{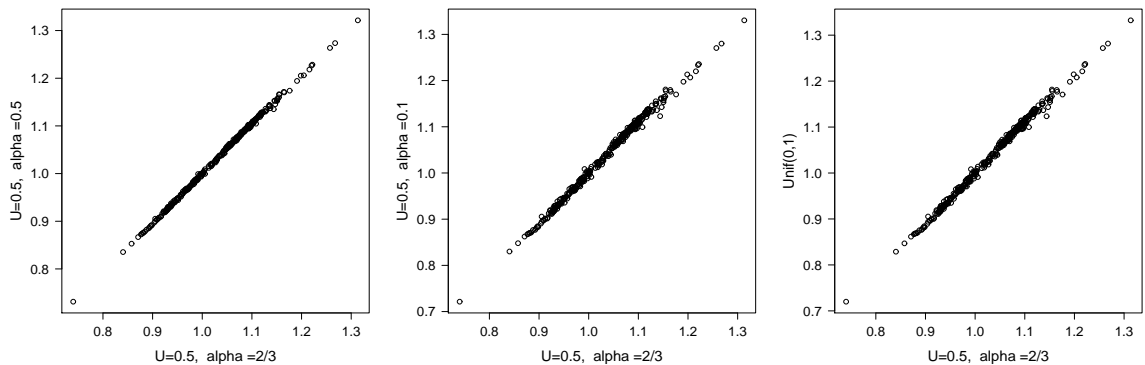}
\caption{Comparison of the posterior mean of the relative risk under
  four different prior distributions for the mixing parameter $\phi$ for IDDM
  counts in Sardinia.}\label{fig:sardiniaRes}
\end{figure}

MacNab et al. \cite{macnab-etal-2004} noted considerable sensitivity
regarding the prior choice for $\phi$ in the Leroux model. For the
Sardinia data, we
empirically check sensitivity by changing the parameters of the
PC-prior for $\phi$ to $U=0.5$ and $\alpha=0.5$, $U=0.5$ and
$\alpha=0.1$. That means a priori
we expect that either $50\%$ or $90\%$ of the variation is explained by the spatially
structured component, respectively. As a fourth alternative we consider
a uniform prior for $\phi$. The rest of Table~\ref{tab:sardiniaEst}
shows the obtained posterior quantities. The estimates for the intercept
stay unchanged. We see a slight change in the posterior estimates for 
$\tau_b$ and $\phi$,
whereby the posterior marginal of $\phi$ still supports almost the whole
range for all priors. The posterior precision slightly decreases,
whereby the $2.5\%$ quantile and the median stay fairly constant. The upper quantile 
and therefore the mean change more clearly. As in MacNab et
al.~\cite{macnab-etal-2004}, the effect on the posterior risk estimates
is negligible (Figure~\ref{fig:sardiniaRes}).

\section{Discussion}\label{sec:discussion}

Throughout this paper we have stressed the importance of scaling in
Bayesian disease mapping. Without proper scaling, hyperparameters have
no clear meaning and may be falsely interpreted. To be more specific,
the precision parameter of the commonly used Leroux and Dean models
tends to be falsely interpreted as marginal precision controlling the
deviation from a constant disease risk over space. However, the parameter depends on the
underlying graph structure and is confounded with the mixing
parameter if the structured spatial component is not appropriately scaled. In addition, it is not clear how to choose a prior
distribution for this parameter. First, due to the lack of scaling, a fixed hyperprior
for the precision parameter gives different amount of smoothing if the graph on which given disease counts are observed is changed. Second,
commonly used hyperprior distributions induce
overfitting\cite{simpson-etal-2015} and will not allow to 
reduce to simpler models such as a constant risk surface or
uncorrelated noise over space. 

Simpson et al. \cite{simpson-etal-2015} proposed a new modification of the commonly known
BYM model, termed BYM2 model,  which addresses the aforementioned issues, and applied it
to one case study.  The BYM2 model
consists of one precision parameter and one mixing parameter. The
precision parameter represents the marginal precision and controls the
variability explained by a spatial effect. The mixing parameter
distributes existing variability between an unstructured and
structured component. Importantly, the structured component is scaled
\cite{sorbye-rue-2014}, which facilitates prior
assignment. Penalised complexity 
priors, i.e.~principle-based priors, are used to favour simpler models until a more complex model is supported. 
Furthermore, these priors allow epidemiologically intuitive specification of
hyperparameters based on the relative risks.

In this paper we have systematically assessed the behaviour of the BYM2 model formulation. 
By means of a simulation study based on the neighbourhood structure of
$366$ regions in Sardinia, we have shown that the model is able to
shrink towards both a constant risk surface or a spatially
unstructured risk for different disease prevalences. This shows that
the model does not overfit. When the disease risk is spatially
structured, the model shows good learning abilities. In contrast to
other model formulations, the
posterior estimates of all hyperparameters can be directly and  intuitively interpreted.
In terms of
model comparison, we have found that the BYM2 model  is
slightly favored compared to the
Leroux and Dean model in terms
of DIC and logarithmic score in the case of a constant risk. In the
case of an unstructured risk the used model choice characteristics are
almost indistinguishable. We think that the practical benefits in terms
of interpretability and prior assignment makes the BYM2 model
advantageous compared to existing models and we recommend its usage
also since its model criteria performance is at least as good as for existing methods.  

All models were
implemented using INLA \cite{rue-etal-2009}, which provides efficient
Bayesian inference without the need of MCMC sampling. This facilitated
the simulation study considerably and allowed us to investigate different
simulation settings. The user-friendly implementation is illustrated
by the R-code in~\ref{app:rcode}. 

MacNab et al.\cite{macnab-etal-2004} noted sensitivity in choosing a prior
distribution for the mixing parameter. Here, we investigated this issue
empirically in both the simulation study and an application to insulin-dependent
diabetes mellitus in Sardinia. However, more effort needs to be placed
into a detailed prior sensitivity analysis following the theoretical
framework of  Roos et al. \cite{roos-etal-2014}.

The BYM2 model can be naturally combined with covariate
information, see Simpson et al\cite{simpson-etal-2015} for an
application, or integrated into a space-time context. However, it will require further work to distribute the variance not only within
the spatial components but over all model parameters in the linear
predictor. It should be noted that the BYM2 model is not only
interesting within disease mapping but also within other fields, such as
genetics, where genes can be regarded as regions and the neighbourhood
structure represents which genes are linked. 

\bibliography{literature.bib}
\bibliographystyle{SageV}

\clearpage

\newcommand{\Appendix}{\appendix\def\thesection{Appendix~\Alph{section}}
\def\thesubsection{\Alph{section}.\arabic{subsection}}
}

\begin{appendix}
\Appendix 

\section{Supplementary tables and figures}\label{app:figtab}

\begin{table}[ht]
\small\sf\centering
\caption{Mean values of the intercept and hyperparameters,
 for the BYM2 model for varying disease
  prevalence, using either a constant, spatially-varying unstructured
  risk surfaces or spatially-varying structured risk surfaces.  A uniform
  prior is used for the weighting parameter $\phi$ while a PC-prior is
  used for $\tau_b$. Standard deviations are
  provided in parentheses.  Results are based on $200$ simulations
for each setting, where the true  parameter values are provided in the table.} \label{tab:simBYM2_unif}

\begin{tabular}{lccc}
  \toprule
  & $\mu$  & $1/\sqrt{\tau}$ & $\phi$ \\
  \midrule
  \multicolumn{4}{l}{Constant risk ($1/\sqrt{\tau_b} = 0$, $\phi=0$):} \\[.2cm]
E = 15 & -7.50E-05 (0.01) & 0.02 (0.02) & 0.46 (0.05) \\ 
  E = 60 & -8.39E-04 (0.01) & 0.01 (0.01) & 0.46 (0.05) \\ 
  E = 200 & -4.15E-04 (0.004) & 0.01 (0.004) & 0.46 (0.04) \\[0.2cm]
  \multicolumn{4}{l}{Area-specific independent risk ($1/\sqrt{\tau_b}
    = 0.5$, $\phi = 0$):}  \\[.2cm]
E = 15 & 1.69E-03 (0.03) & 0.50 (0.03) & 0.07 (0.04) \\ 
  E = 60 & -1.91E-03 (0.03) & 0.50 (0.02) & 0.06 (0.03) \\ 
  E = 200 & 1.21E-03 (0.03) & 0.51 (0.03) & 0.06 (0.03) \\[0.2cm]
  \multicolumn{4}{l}{Area-specific correlated risk ($1/\sqrt{\tau_b}
    = 0.5$, $\phi = 1$):}  \\[.2cm]
E = 15 & 2.15E-03 (0.01) & 0.47 (0.03) & 0.90 (0.05) \\ 
  E = 60 & 1.32E-05 (0.01) & 0.49 (0.06) & 0.93 (0.04) \\ 
  E = 200 & 4.83E-05 (0.004) & 0.49 (0.07) & 0.93 (0.03) \\ 
   \bottomrule
 \end{tabular}
\end{table}

\begin{figure}[h]
\centering
\includegraphics[width=.95\textwidth]{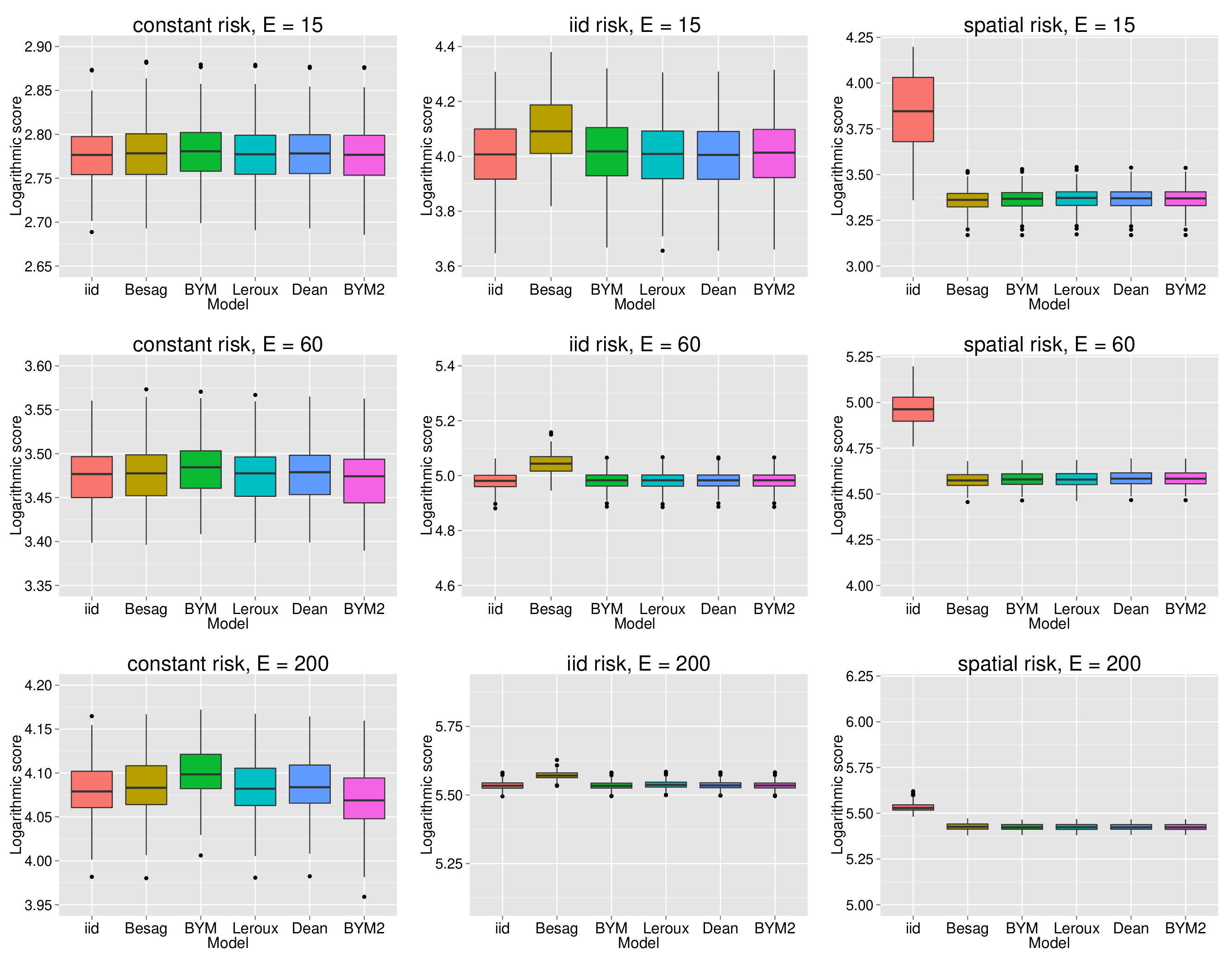}
\caption{Boxplot of logarithmic scores for all six models for nine different simulation
  settings.  For each setting 200 simulations were
  generated. The left column shows results when simulating a constant
  risk, the middle column those for an iid risk surface and the right
  column those for a spatially structured risk surface. The rows
  represent varying disease prevalence with $E=\{15,60,200\}$. Please
  note that the y-axis range width is only kept constant
  column-wise. For ease of visibility, the boxplots of the iid-model
  for $E=15$ and $E=60$ in  the right column are slightly trimmed from
  the top.
}\label{fig:cpo}
\end{figure}

\begin{figure}
\centering
\includegraphics[width=.95\textwidth]{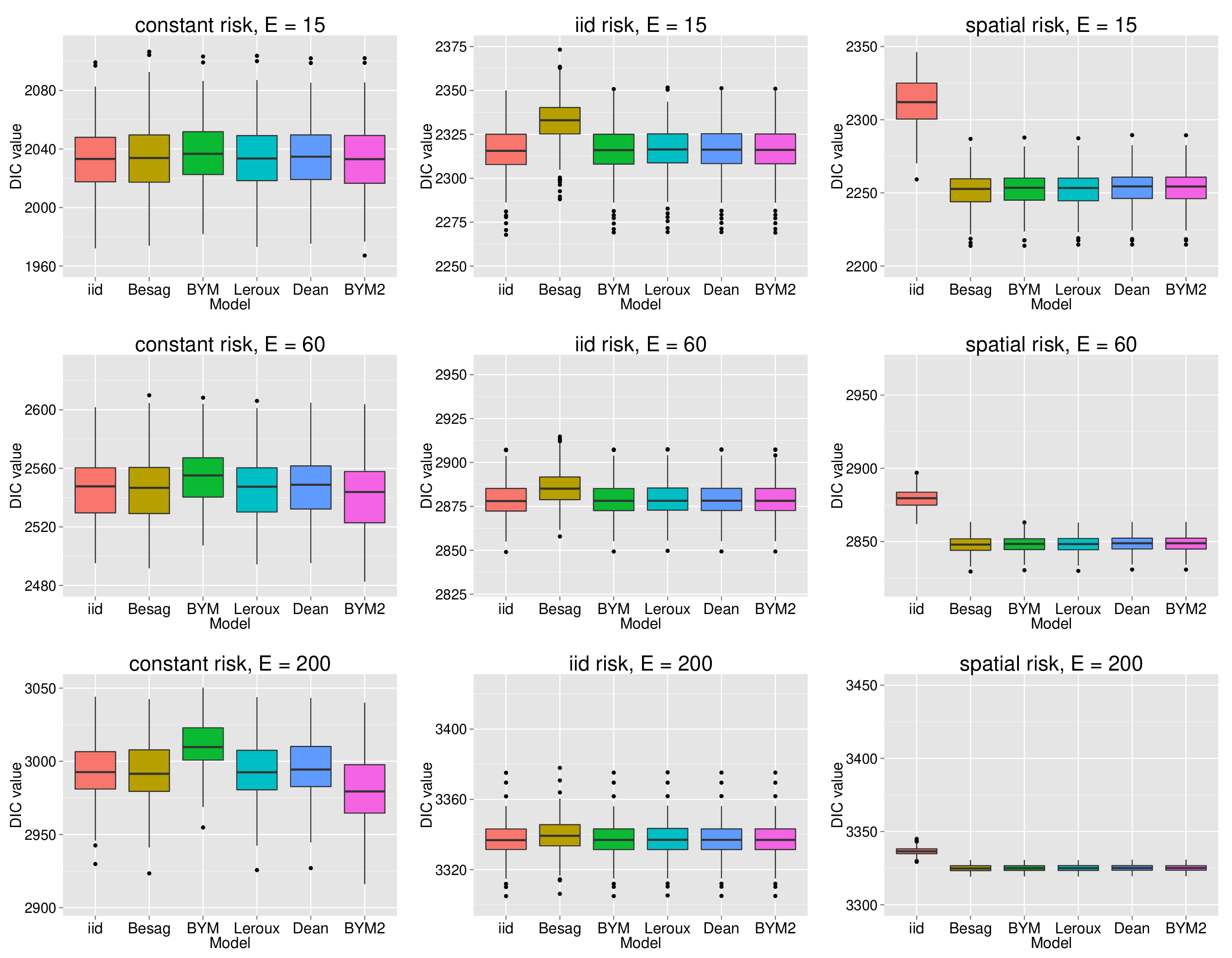}
\caption{Boxplot of DIC values for all six models for nine different simulation
  settings.  For each setting 200 simulations were
  generated. The left column shows results when simulating a constant
  risk, the middle column those for an iid risk surface and the right
  column those for a spatially structured risk surface. The rows
  represent varying disease prevalence with $E=\{15,60,200\}$. Please
  note that the y-axis range width is only kept constant column-wise. 
For ease of visibility, the boxplot of the iid-model
  for $E=15$  in  the right column is slightly trimmed from
  the top.
}\label{fig:dic}
\end{figure}

\begin{table}
\small\sf\centering
\caption{Root mean square error (RMSE), DIC and logarithmic score
  (LC) when using different models for analysing IDDM counts in
  Sardinia. Four different priors for the mixing parameter $\phi$ are
  compared: PC 1 - $U=0.5, \alpha=2/3$, PC 2 -  $U=0.5, \alpha=0.5$, 
  PC 3 - $U=0.5, \alpha=0.1$, uniform on $(0,1)$.}\label{tab:mc-sardinia}
\begin{tabular}{rrrr}
  \toprule
 Model & RMSE & DIC & LS \\ 
  \midrule
iid-model & 1.35 & 838 & 1.15 \\ 
  Besag & 1.35 & 831 & 1.14 \\ 
  BYM & 1.34 & 832 & 1.14 \\ 
  Leroux & 1.34 & 835 & 1.14 \\ 
  Dean & 1.34 & 834 & 1.14 \\ 
  BYM2 (PC1) & 1.33 & 834 & 1.14 \\ 
  BYM2 (PC2) & 1.34 & 833 & 1.14 \\ 
  BYM2 (PC3) & 1.34 & 833 & 1.14 \\ 
  BYM2 (Unif) & 1.34 & 833 & 1.14 \\ 

    \bottomrule
\end{tabular}
\end{table}

\clearpage

\section{INLA R-code to implement the BYM2 model} \label{app:rcode}

In the following, we illustrate how the BYM2 model can be implemented
within the R-package {\tt INLA}. 

\small
\begin{lstlisting}
# load the R-package
library(INLA)

# provide the path to the file storing the neighbourhood structure
g = "sardinia.graph"
# read the data
sardinia = read.table("sardinia.dat", col.names=c("y", "E", "SMR"))

# show the first lines
head(sardinia)
#   y          E       SMR
#1  1  0.5986411 1.6704500
#2  0  0.6055964 0.0000000
#3 10 13.3658700 0.7481743
#4  0  0.2346664 0.0000000

# get the number of regions
n = nrow(sardinia)
# define a region index with values 1, 2, ..., n
sardinia$region = 1:n

# specify the latent structure using a formula object
formula = y ~ 1 + f(region,
                    model = "bym2",
                    graph=g,
                    scale.model = TRUE,
                    constr = TRUE,
                    hyper=list(
                      phi = list(
                        prior = "pc",
                        param = c(0.5, 2/3),
                        initial = -3),
                      prec = list(
                        prior = "pc.prec",
                        param = c(0.2/0.31, 0.01),
                        initial = 5)))

# call the inla function
result = inla(formula, data = sardinia, family = "poisson", E=E,
         control.predictor = list(compute=TRUE))
# get improved estimates for the hyperparameters
result = inla.hyperpar(r, dz = 0.2, diff.logdens=20)
\end{lstlisting}

First we have to load the INLA package which can be installed by
typing 
\begin{lstlisting}
install.packages("INLA", repos="http://www.math.ntnu.no/inla/R/testing")
\end{lstlisting}
in the R-terminal. In this paper we used the INLA version
0.0-1445883880. In order to define the model incorporating a spatially
structured component we need to provide the neighbourhood structure
for the regions we analyse. In INLA this is done by providing the
path to the file which stores this information. The graph file for
Sardinia is named {\tt sardinia.graph} and has the following structure:
\begin{verbatim}
366
0 5 13 61 69 73 81
1 5 8 16 40 46 94
2 4 47 59 63 77
3 3 11 17 43
4 5 24 41 51 56 67
...
363 5 292 307 324 344 346
364 5 291 310 317 329 345
365 3 289 303 308
\end{verbatim}

The first line provides the number of regions, here $366$. The following
$366$ lines, corresponds to each of the regions, where the first entry always
specifies a unique region index. The second entry specifies the
number of neighbouring regions, and then the corresponding region
indices for those neighbours are listed. INLA transforms this
information into the neighbourhood matrix $\mathbf{Q}$ with entries as 
specified in \eqref{eq:besagQ}. 

Lines 23--36 specify
the model structure in terms of a formula object. Here, we have an
intercept and a latent Gaussian model given by  \eqref{eq:bym2}. The
{\tt f(.)} function is used to specify a random effect in {\tt
  INLA}. The first argument is always the variable name to which it applies,
here region. We specify the use of the BYM2 model in line 24. Then, we
provide the graph and specify that we want to scale the model in line
26. This will scale the graph as described in Section
\ref{sec:scaling}. To avoid confounding with the intercept a
sum-to-zero constraint is specified in line 27. The next lines specify
the hyperprior for the two hyperparameters in form of a list of length
two. The first list entry specifies the prior for the mixing parameter
$\phi$, named {\tt phi} in the INLA implementation, the second entry for the precision parameter 
$\tau_b$, named {\tt prec}. Each entry is itself a list. Using the
argument {\tt prior} we specify the prior we would like to use. The
parameters are set in the argument {\tt param}. They coincide with the
values used in Section~\ref{sec:sardinia}. Using the argument {\tt
  initial} a starting value for the numerical optimisation of the INLA
algorithm can be provided. This has to be given on the log scale for
the precision and the logit scale for $\phi$, as these are the scales INLA
works on internally.

After defining the model, we call the {\tt inla} function providing the
{\tt formula}, the data and the likelihood. Further, we specify the
expected number of counts and specify that we not only would like to
get posterior estimates for the components of $\eta_i$ but also
$\eta_i$ and $\theta_i$ (line 40). Improved hyperparameter estimates
can be obtained by calling {\tt inla.hyperpar} afterwards. Here, {\tt
  dz} and {\tt diff.logdens} refer to options in the numerical
integration algorithm, see help page. 

Results can be inspected using {\tt summary(result)} or {\tt
  plot(result)}. Posterior marginals for the fixed effects and hyperparameters can be
consequently accessed via {\tt result\$summary.fixed} or {\tt
  result\$summary.hyperpar}, respectively.

\section{PC prior derivation}\label{app:pc}

\subsection{Precision parameter $\tau_b$}\label{app:pc-prec}
In deriving the PC prior for $\tau_b$, we start by assuming a large fixed precision $\tau_0$
for the base model, where $\tau_0 \gg \tau_b$. 
 The resulting $n-1$-dimensional stochastic part of the Kullback-Leibler divergence is
\begin{gather}
\text{KLD}(\tau_b) = \frac{n-1}{2}\frac{\tau_0}{\tau_b}
\left
(1+\frac{\tau_b}{\tau_0} \log \left[\frac{\tau_b}{\tau_0}\right] -
\frac{\tau_b}{\tau_0}\right)
\rightarrow \frac{n-1}{2}\frac{\tau_0}{\tau_b}, 
\end{gather}
which implies
$d(\tau_b) \rightarrow \sqrt{(n-1)\tau_0/\tau_b}$. 

The next step is to assign a prior to the distance $d(\tau_b)$ and transform this  by a change of variables. 
We use the exponential prior with decay rate $\lambda$, which
supports the constant-rate penalisation and apply the ordinary change of variable transformation
\begin{gather*}
  \pi(\tau_b) = \lambda \exp(-\lambda d(\tau_b)) \left|
    \frac{\partial d(\tau_b)}{\partial \tau_b}\right|. 
\end{gather*}
Formulating this expression out and setting $\theta = \lambda\sqrt{(n-1)\tau_0}$,  where $\lambda$ is
chosen so that $\theta$ is kept constant in the limit $\tau_0
\rightarrow \infty$, we obtain a 
type-2 Gumbel distribution\citep{simpson-etal-2015}
\begin{gather*}
  \pi(\tau_b) = \frac{\theta}{2} \tau_b^{-3/2} \exp\left(-\theta
    \tau_b^{-1/2}\right).
\end{gather*}
as prior for $\tau_b$.

\subsection{Mixing parameter $\phi$}\label{app-mixing}
To derive the PC prior for $\phi$, we assume that  $\mathbf{Q}_\star \geq 0$
and define the covariance matrices for the base and flexible models as
$\mathbf{\Sigma}_{\text{base}} =  \mathbf{I}$ and
$\mathbf{\Sigma}_{\text{flex}}(\phi) =  (1-\phi) \: \mathbf{I} + \phi\mathbf{Q}_\star^{-}$, respectively. The  resulting Kullback-Leibler divergence is
\begin{align*}
  \text{KLD}(\mathcal{N}(0,\mathbf{\Sigma}_{\text{flex}}(\phi)) \mid
  \mathcal{N}(0, \mathbf{\Sigma}_{\text{base}})) &=
                                                   \frac{1}{2}(\text{trace}(\mathbf{\Sigma}_{\text{flex}}(\phi))
  - n - \log|\mathbf{\Sigma}_{\text{flex}}(\phi)|)\\
&= n\phi \left(\frac{1}{n}\text{trace}( \mathbf{Q}_\star^{-1}) - \log
  |(1-\phi) \mathbf{I} + \phi\mathbf{Q}_\star^{-1} |\right).
\end{align*}
The computation of the trace is quick when $\mathbf{Q}_\star$ is
sparse \cite{erisman-tinney}. The determinant can be computed as $|(1-\phi) \mathbf{I} + \phi\mathbf{Q}_\star^{-1} | =
\prod_{i=1}^n(1-\phi + \phi\tilde{\gamma}_i)$,
 where $\{\gamma_i\}$ denote the eigenvalues of  $\mathbf{Q}_\star$, and $\tilde{\gamma}_i
= 1/\gamma_i$  \cite{simpson-etal-2015}. As the given precision matrix
$\mathbf{Q}_\star$ is singular with rank deficiency 1, one of the
eigenvalues is zero. This implies that  $\tilde{\gamma}_i
= 1/\gamma_i$ if $\gamma_i > 0$,  or else $\tilde{\gamma}_i = 0$. We
can avoid calculating eigenvalues if the dimension is high, see
Appendix A3 of Simpson et al. \cite{simpson-etal-2015} for details.
\end{appendix}

\end{document}